\documentclass[twocolumn,showpacs,preprintnumbers,draftclsnofoot]{revtex4}
\usepackage{amsmath}
\usepackage{amssymb}
\usepackage{graphicx}
\usepackage{dcolumn}
\usepackage{bm}
\usepackage{amssymb}
\usepackage{graphicx}
\usepackage{dcolumn}
\usepackage{bm}

\begin{document}

\title{ Antiferromagnetic spin valve from heterostructure of two-dimensional hexagonal crystals }
\author{Ma Luo\footnote{Corresponding author:luom28@mail.sysu.edu.cn} }
\affiliation{The State Key Laboratory of Optoelectronic Materials and Technologies \\
School of Physics\\
Sun Yat-Sen University, Guangzhou, 510275, P.R. China}

\begin{abstract}

Spin valves consisting of heterostructures of single-layer hexagonal crystal on an antiferromagnetic substrate or of bilayer hexagonal crystal intercalated between two (anti)ferromagnetic insulators, with the current-in-plane geometry, are proposed. The two-dimensional hexagonal crystals such as graphene, silicene, germanene, and stanene are modeled by the tight binding model of honeycomb lattice. The magnetization orientation of the antiferromagnetic substrate(s) controls the band gap and topological properties of bulk, which in turn control the transport of three types of spin valve geometries: (i) the in-plane transport of bulk; (ii) the transport of topological edge states along nanoribbon with bulk gap; (iii) the transport of chiral edge state along domain wall. The heterostructures are investigated by a tight binding model with an (anti)ferromagnetic exchange field, Hubbard interaction and(or) spin-orbital coupling. For the first type of spin valve geometry, the Hubbard interaction could enlarged the effective band gap of bulk, which in turn improve the sensitivity of the spin valves to the antiferromagnetic exchange field. For the second and third types of spin valve geometries, the topological phase diagrams of varying types of heterostructures with spin-orbital coupling serve as guideline for designing the spin valve. The coexistence of the Hubbard interaction and the spin-orbital coupling could enlarge the topological gap in bulk and improve the quality of the chiral edge states at the domain walls between regions with different topological numbers.

\end{abstract}

\pacs{00.00.00, 00.00.00, 00.00.00, 00.00.00} \maketitle

\section{Introduction}

Spintronics has been proposed as a physical foundation for information processing and storage systems. The phenomenon of spin-dependent transport based on the physics of giant magnetoresistance \cite{Baibich88,Binasch89} and
tunnel magnetoresistance \cite{Julliere75,Moodera95} have been intensively studied, which led to the concept of the spin valve \cite{Dieny91}, which controls electronic conductivity by magnetization orientation. To reduce the power consumption and device size of spintronic systems, antiferromagnetic materials have been proposed to replace the ferromagnets \cite{Jungwirth16,Baltz18}. Spintronic devices based on antiferromagnetic materials have multiple advantages over ferromagnetic spintronic devices \cite{Zelezny18}, such as the absence of parasitic stray fields and ultrafast magnetization dynamics. Therefore, recent efforts have been devoted to developing spin valves based on antiferromagnetic materials \cite{BGPark11}. Antiferromagnetic insulators with high Neel temperature have been experimentally synthesized \cite{Belashchenko10,Prakash16,Kampfrath11,Bhattacharya16}, including materials with hexagonal crystal structure \cite{Belashchenko10,Prakash16}.

Two-dimensional (2D) hexagonal crystals such as graphene, silicene, germanene, and stanene has been considered as a preeminent candidate for spintronic systems \cite{Pesin12,WeiHan14,Kheirabadi14,YangWang15}. Graphene has high electron mobility \cite{Avouris07,CastroNeto09} and long spin relaxation lifetime \cite{Min06,Hernando06,Boettger07,Garcia18}. The generation and manipulation of spin current by electronic \cite{Grujic14,DaPingLiu16} and optical \cite{Inglot14,Rioux14,Inglot15,Kaladzhyan15,luo17,Reinaldo17} methods have been proposed theoretically and studied experimentally \cite{Mendes15,Gurram16,Omar17}. Single-layer graphenes (SLGs) and bilayer graphenes (BLGs) with the substrate proximity effect\cite{Rashba09,Zhenhua10,Jayakumar14} or adatom doping\cite{Fufang07} exhibit large values of spin-orbital coupling (SOC) that make the spintronic effect observable at room temperature. Specifically, in systems of graphene on transition-metal dichalcogenides (TMDCs), staggered and Rashba SOC is introduced in the graphene, resulting in novel spin effects and topological properties \cite{Gmitra15,Gmitra16,Morpurgo15,Offidani17,Frank18,Abdulrhman18}. The proximity effect of graphene on the ferromagnetic insulator induces a ferromagnetic exchange field \cite{VoTienPhong18}. The coexistence of the ferromagnetic exchange field and Rashba SOC drive the SLGs into quantum anomalous Hall (QAH) phase \cite{ZhenhuaQiao10,MotohikoEzawa12,Offidani18}. The topological zero-line mode at the domain wall of the QAH phase host topological edge states with localized one-way charge transportation \cite{Yafei17}. Moreover, van der Waals spin valves based on BLG heterostructures with ferromagnetic insulators \cite{YuanLiu16} have been recently proposed \cite{Cardoso18}. The presence of Hubbard interaction in the graphene model induces spontaneous antiferromagnetic order at the zigzag edge \cite{Rossier08,Jung09,Lado14} and has been proposed for use as a spin injector \cite{Soriano12} or spin valve \cite{Friedman17}. On the other hand, silicene have larger intrinsic SOC and tunable staggered sublattice potential \cite{Alessandro17}, which could be designed as varying type of spin current generator \cite{Motohiko12,XiaoLong18}.

To harness the advantages of both 2D hexagonal crystals and antiferromagnetic spintronics, we propose spin valves based on heterostructures containing 2D hexagonal crystals and antiferromagnetic materials as substrates. The spin valves have current-in-plane (CIP) geometry, i.e. the magnetization orientation of the antiferromagnetic substrate control the in-plane conductivity (conductance) of the conducting layer in bulk (nanoribbon). The conducting layer, which could be single layer or bilayer graphene (or silicene), is described by the tight binding model of honeycomb lattice with staggered sublattice exchange field. Such heterostructures have recently been theoretically \cite{XiaoLong18,LeiXu18} and experimentally \cite{YanFeiWu17,BoyiZhou18} studied. Assuming that an antiferromagnetic insulator with a surface lattice matched to graphene could be found or engineered, a staggered sublattice exchange field would be induced in the graphene. The staggered sublattice exchange field in silicene could be realized by sandwiching the silicene between two perovskite layers \cite{QiFengLiang13}. We investigated single layer honeycomb lattice models with an antiferromagnetic exchange field, bilayer honeycomb lattice models with antiferromagnetic exchange fields in both layers, and bilayer honeycomb lattice models with a ferromagnetic(antiferromagnetic) exchange field in the top (bottom) layer. We applied the parameters for graphene(SLGs and BLGs) in the models. If the parameters for silicene are used instead, the conclusions would be qualitatively the same.

Depending on the configuration of the conducting layer and substrates, three types of spin valve in CIP geometries are proposed. (i) Spin valve-I: both conducting layer and antiferromagnetic substrates are uniform bulk. The band gaps of the conducting layer could be controlled by rotating the magnetization orientation of the antiferromagnetic exchange field. The effect of realistic Hubbard interaction and SOC on the band gaps has been examined. (ii) Spin valve-II: the conducting medium is a zigzag nanoribbon, and the antiferromagnetic substrates are uniform bulk. The zigzag nanoribbon has bulk gap. The conductance of the zigzag nanoribbon is originated from the zigzag edge states. The topological properties of the corresponding bulk determined the properties of the zigzag edge states. Rotation of the magnetization orientation of the antiferromagnetic exchange field changes the topological properties, which in turn control the presence or absence of the conducting zigzag edge states. Topological phase diagrams of varying types of heterostructures are discussed. Some phase regimes have a nonzero Chern number but a small or vanishing global band gap. For bilayer honeycomb lattice model, the coexistence of SOC and Hubbard interaction modifies the phase boundaries and enlarges the band gap, stabilizing the QAH phase in these regimes. (iii) Spin valve-III: the conducting layer is zigzag nanoribbon with large width, and the substrates are bulk with antiferromagnetic domain wall along the axis of the nanoribbon. A kink of antiferromagnetic exchange field is induced in the honeycomb lattice model. At the two sides of the kink, the bulk band gaps are in the same energy range, and the topological properties are different. Thus, the kink hosts gapless localized chiral edge states. The conductance of the zigzag nanoribbon is originated from the chiral edge states. Rotation of the magnetization orientation of the antiferromagnetic exchange field in one half of the substrate controls the presence or absence of the kink, which in turn control the presence or absence of the conducting chiral edge states.

For the single layer honeycomb lattice models, spin valve-I and spin valve-III are proposed. For the bilayer honeycomb lattice models, all three type of spin valves are proposed. The article is organized as follows: In section II, single layer honeycomb lattice models with antiferromagnetic exchange fields are studied. The effects of Hubbard interaction to bulk and zigzag nanoribbons are separately discussed in subsections A and B, respectively. The effect of SOC to bulk is discussed in subsection C. In section III, bilayer honeycomb lattice models with (anti)ferromagnetic exchange field are studied. The effects of Hubbard interaction and SOC are separately discussed in subsections A and B, respectively. The effect of the coexistence of Hubbard interaction and SOC is discussed in subsection C. The chiral edge states of the domain wall are discussed in subsection D. Section IV presents the conclusion.

\section{single-layer honeycomb lattice}


The Hamiltonian of the single-layer honeycomb lattice model with antiferromagnetic exchange field is given as
\begin{equation}
H=H_{0}+H_{AF} \label{HamiltonianSLGAF}
\end{equation}
where
\begin{equation}
H_{0}=-\sum_{\langle i,j\rangle,\sigma}tc_{i\sigma}^{+}c_{j\sigma} \label{Hamiltonian0}
\end{equation}
\begin{eqnarray}
H_{AF}&=&\lambda_{AF}^{x}\sum_{i,\sigma,\sigma'}\kappa_{i}\hat{s}^{x}_{\sigma,\sigma'}c_{i\sigma}^{+}c_{i\sigma'}\nonumber \\
&+&\lambda_{AF}^{y}\sum_{i,\sigma,\sigma'}\kappa_{i}\hat{s}^{y}_{\sigma,\sigma'}c_{i\sigma}^{+}c_{i\sigma'}\nonumber\\
&+&\lambda_{AF}^{z}\sum_{i,\sigma,\sigma'}\kappa_{i}\hat{s}^{z}_{\sigma,\sigma'}c_{i\sigma}^{+}c_{i\sigma'}
\label{HamiltonianAF}
\end{eqnarray}
The Hamiltonian of a pristine SLG is $H_{0}$, where $t=2.8$ eV is the hopping energy, $\sigma=\pm1$ is the index of spin, $i$ and $j$ are the indices of lattice sites, and $c_{i\sigma}^{+}$($c_{i\sigma}$) is the creation (annihilation) operator of the electron at the $i$-th lattice site with spin index $\sigma$. The summation indices with $\langle i,j\rangle$ cover the nearest neighboring lattice sites. The antiferromagnetic exchange field along $x$, $y$ and $z$ direction is modeled by a spin dependent staggered sublattice potential. In Eq. (\ref{HamiltonianAF}), $\kappa_{i}$ is equal to $1$($-1$) for the A(B) sublattice, $\hat{s}^{x,y,z}$ are the Pauli matrix of spin $x$, $y$ and $z$, and $\lambda_{AF}^{x,y,z}$ is the strength of the antiferromagnetic exchange field. For silicene being intercalated between perovskite layers, the magnitude of the antiferromagnetic exchange field, which is designated as $\lambda_{AF}=\sqrt{(\lambda_{AF}^{x})^{2}+(\lambda_{AF}^{y})^{2}+(\lambda_{AF}^{z})^{2}}$, could be given by DFT calculations \cite{QiFengLiang13}. The DFT calculations of graphene proximity-coupled to an antiferromagnetic insulator with large lattice mismatching \cite{ZhenhuaQiao1411} show that only a ferromagnetic exchange field is effectively induced. An antiferromagnetic substrate with small lattice mismatching would induce antiferromagnetic exchange field with a spatial $Moir\acute{e}$ pattern or the combination of ferromagnetic and antiferromagnetic exchange field. In this article, we assumed sole presence of antiferromagnetic or ferromagnetic exchange field in the honeycomb lattice model. The model Hamiltonian (\ref{HamiltonianSLGAF}) can also describe functionalized tin films X-Sn (where X= H, F, Cl, Br, and I) \cite{CNiu17} with honeycomb lattice structure.

The band structure of Eq. (\ref{HamiltonianSLGAF}) in bulk exhibit massive Dirac Fermion with gap $2\lambda_{AF}$. As comparison, the nonmagnetic model with Hamiltonian
\begin{equation}
H=H_{0}+H_{\Delta} \label{HamiltonianSLG_D}
\end{equation}
where
\begin{equation}
H_{\Delta}=\lambda_{\Delta}\sum_{i,\sigma}\kappa_{i}c_{i\sigma}^{+}c_{i\sigma}
\label{HamiltonianD}
\end{equation}
also exhibit massive Dirac Fermion with gap $2\lambda_{\Delta}$. However, the topological properties of the model Hamiltonian (\ref{HamiltonianSLGAF}) and (\ref{HamiltonianSLG_D}) are different. The topological properties of the bulk, such as the Chern number, valley Chern number and spin-dependent valley Chern number, can be calculated from the band structure. The Chern number $\mathcal{C}$ is the integral of the Berry curvature through the whole first Brillouin zone \cite{ZhenhuaQiao10,MotohikoEzawa12}. Because the Berry curvature value is large in the vicinity of the $K$($K^{\prime}$) point, the Chern number of each valley, designated as $\mathcal{C}_{K(K^{\prime})}$, can be defined for the corresponding continuum Dirac fermion model \cite{FanZhang13,ZhenhuaQiao13}. The difference between the Chern number of $K$ and the $K^{\prime}$ valley is the valley Chern number $\mathcal{C}_{V}=\mathcal{C}_{K}-\mathcal{C}_{K^{\prime}}$. For the Hamiltonian with no spin flipping term, the Chern number of each spin component and each valley can be defined for the corresponding continuum Dirac fermion model as well, designated as $\mathcal{C}_{K(K^{\prime})}^{\sigma}$. The spin-dependent valley Chern number is defined as $\mathcal{C}_{V}^{\sigma}=\mathcal{C}_{K}^{\sigma}-\mathcal{C}_{K^{\prime}}^{\sigma}$. The calculation results show that the spin- and valley-dependent Chern number of the Hamiltonian (\ref{HamiltonianSLGAF}) is $C_{\tau}^{\sigma}=\tau\sigma$ with $\tau=\pm1$ labeling the $K$($K^{\prime}$) valley. As comparison, the valley Chern number of the Hamilsonian (\ref{HamiltonianSLG_D}) does not depend on spin. The domain wall along zigzag direction between two regions with staggered sublattice potentials of opposite sign \cite{Semenoff08,Zarenia12} host chiral edge states. The velocities of the chiral edge states at opposite valleys are opposite to each other. For the geometry of spin valve-III, the domain wall separate two regions with opposite antiferromagnetic exchange field. The chiral edge states of opposite spins have opposite velocities. As a result, the chiral edge states support dissipationless spin-valley current at the intrinsic Fermi level \cite{KyuWonLee17}.

\subsection{Bulk band gap: effect of Hubbard interaction}

In the presence of Hubbard interaction, an addition term $H_{U}$ is added to the Hamiltonian (\ref{HamiltonianSLGAF}) and (\ref{HamiltonianSLG_D}). The Hubbard model is given as
\begin{equation}
H_{U}=U\sum_{i}n_{i\sigma}n_{i\bar{\sigma}}
\label{HamiltonianU}
\end{equation}
where $n_{i\sigma}$ is the operator of the particle number at the $i$-th lattice with spin indices $\sigma$ and $\bar{\sigma}=-\sigma$. For realistic graphene(or silicene), the interaction parameter is $U=1.6t$ \cite{Schuler13}, which is used in the rest of this article. For the interacting model, the single particle band structure is not well defined. Instead, the single particle Green's function describes the quasi-particle dynamic. The corresponding spectral function describes the distribution of quasi-particle in the energy-momentum phase space. For bulk, the Green's function is a function of Bloch wave vector, $\mathbf{k}$, and energy, $\varepsilon$, which can be formally written as
\begin{equation}
G(\mathbf{k},z)=[z-H_{free}(\mathbf{k})-\Sigma(\mathbf{k},z)]^{-1}
\label{greenfunction}
\end{equation}
where $z=\varepsilon+i0$ is the energy with positive infinitesimal imaginary part, $H_{free}(\mathbf{k})$ is the non-interacting part of the Hamiltonian with Bloch phase, and $\Sigma(\mathbf{k},z)$ is the self-energy. Both $G(\mathbf{k},z)$ and $\Sigma(\mathbf{k},z)$ are matrix with size being equal to two times of the number of lattice sites within one primitive unit cell. For the strongly correlated model with $U>t$, cluster perturbation theory (CPT) \cite{Pairault98,Senechal00,Senechal02,Grandi15,Grandi151,luo181} is an efficient method to calculate the Green's function. Short-range dynamical correlations as well as on-site correlations are captured by CPT. The details of calculating the Green's function by CPT method is presented in appendix. The spectral function is defined as $A(\mathbf{k},\varepsilon)=-\frac{1}{\pi}Im\{tr[G(\mathbf{k},z)]\}$. The band gap is extracted from $A(\mathbf{k},\varepsilon)$.



\begin{figure}[tbp]
\scalebox{0.58}{\includegraphics{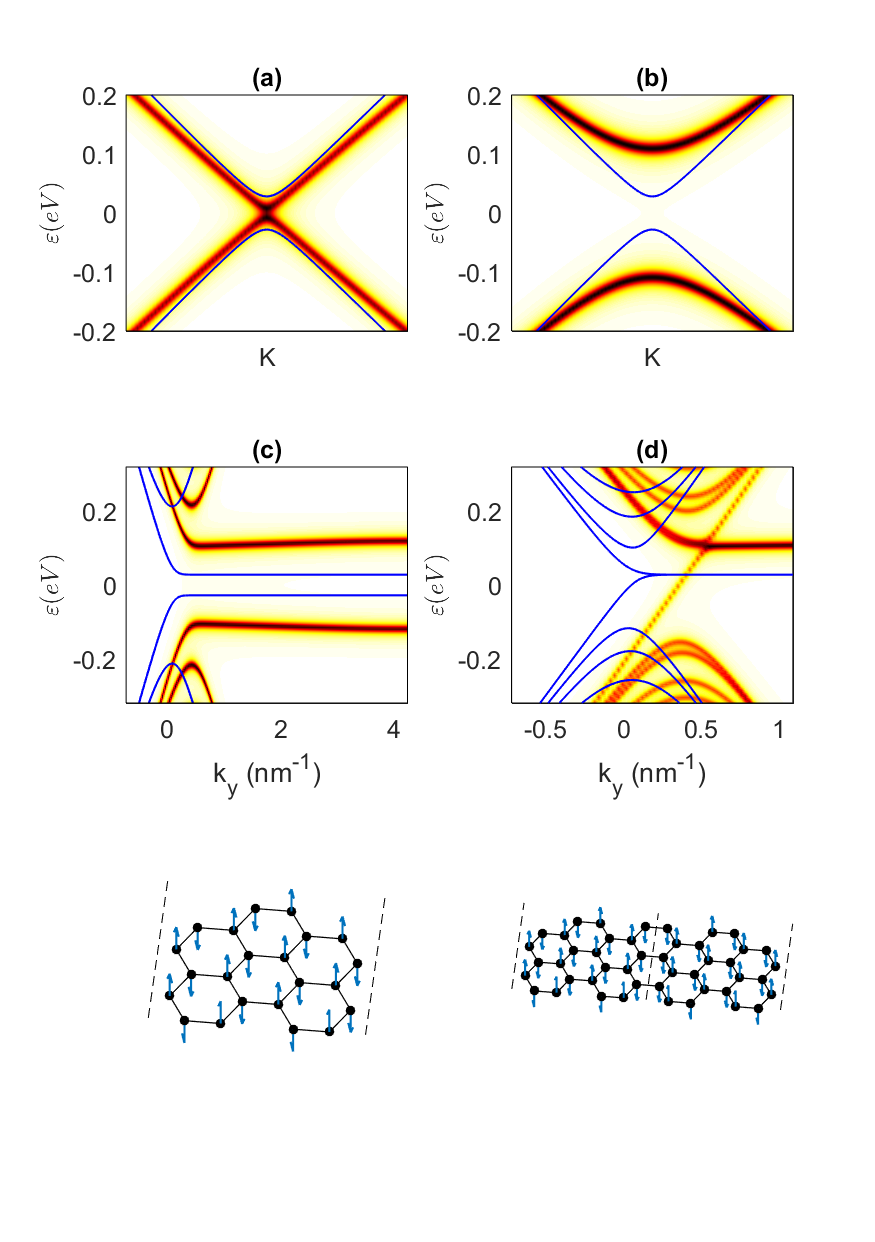}}
\caption{ The shaded surface and blue solid lines are the spectral function and band structure, respectively. Figure (a) represents the Hamiltonian $H_{0}+H_{\Delta}+H_{U}$ in the bulk system. Figures (b), (c) and (d) represent the Hamiltonian $H_{0}+H_{AF}+H_{U}$ ($\lambda^{x}_{AF}=\lambda^{y}_{AF}=0$ and $\lambda^{z}_{AF}=0.01t$) in bulk, zigzag nanoribbon (60 unit cells) and zigzag nanoribbon (120 unit cells) with a domain wall, respectively. The magnetized orientation of the exchange fields of the nanoribbons in (c) and (d) are shown by the arrows in the pictures below each figure, with dashed lines marking the open boundaries and the domain wall. Only the band structure and spectral function of the spin-up component are plotted in (d).   }
\label{fig_SLGgap}
\end{figure}

The two models with Hamiltonian in Eq. (\ref{HamiltonianSLGAF}) and (\ref{HamiltonianSLG_D}) respond to the Hubbard interaction differently. To compare the two models, we perform calculation by CPT, assuming the parameters $\lambda_{AF}=0.01t$($\lambda^{x}_{AF}=\lambda^{y}_{AF}=0$ and $\lambda^{z}_{AF}=0.01t$) and $\lambda_{\Delta}=0.01t$ in the former and latter model, respectively. The noninteracting band structures of the two models are the same, as shown in Fig. \ref{fig_SLGgap}(a) and (b). In the presence of interaction, the spectral function given by CPT is plotted as a shaded surface in Fig. \ref{fig_SLGgap}(a) and (b). For the Hamiltonian $H_{0}+H_{\Delta}+H_{U}$, the gap of the Dirac fermion is suppressed to near zero, as shown in Fig. \ref{fig_SLGgap}(a). A similar conclusion was given by the theoretical calculation of the Hubbard model with dynamical mean field theory \cite{JinRongXu161,JinRongXu162}. In contrast, for the Hamiltonian $H_{0}+H_{AF}+H_{U}$, the gap is enlarged, as shown in Fig. \ref{fig_SLGgap}(b). This phenomenon can be understood by inspecting the self-energy of the two models. The diagonal elements of the self-energy at the $i$-th site equate approximately to constants, $\Sigma_{i\sigma,i\sigma}\propto U\langle n_{i\bar{\sigma}}\rangle$. The $\langle n_{i\sigma}\rangle$ at the A sublattice is larger (smaller) than that at the B sublattice if the local potential of the opposite spin at the A sublattice is smaller (larger) than that at the B sublattice. The self-energy is approximately equivalent to another staggered sublattice potential. For the models with $H_{0}+H_{\Delta}+H_{U}$, the self-energy partially cancels out the staggered sublattice potential. In contrast, for the models with $H_{0}+H_{AF}+H_{U}$, the self-energy has the same sign as the spin dependent staggered sublattice potential, and thus, the gap is enlarged. Because the self-energy contains nonzero nondiagonal elements, which effectively change the nearest neighbor hopping energy, the Fermi velocities of the interacting systems (the slopes extracted from the spectral function) are smaller than those of noninteracting systems. Because the band structure and spectral function is independent on the direction of the antiferromagnetic exchange field, the single layer honeycomb lattice model of Eq. (\ref{HamiltonianSLGAF}) in bulk does not function as spin valve-I.

\subsection{Zigzag nanoribbon}

A zigzag nanoribbon of the model Hamiltonian (\ref{HamiltonianSLGAF}) has a band gap between two flat bands, as shown in Fig. \ref{fig_SLGgap}(c). In the presence of Hubbard interaction, the flat bands become slightly dispersive. Because the band structure and spectral function is independent on the direction of the antiferromagnetic exchange field, the systems do not function as spin valve-II.

Reversing the antiferromagnetic exchange field in one half of the nanoribbon, as shown in Fig. \ref{fig_SLGgap}(d), creates a domain wall in the middle of the nanoribbon, which hosts gapless spin-dependent chiral edge states. The system functions as spin valve-III. For a noninteracting model in an infinitely wide nanoribbon, the dispersion of the chiral edge state is $\varepsilon=\sigma\tau\hbar v_{F}k_{y}$, where $k_{y}$ is the wavenumber along the domain wall. Because of the finite size effect, the chiral edge state at the domain wall is mixed with the zigzag edge states at the open boundaries, so that the dispersion deviates from linear, as shown by the blue (solid) line in Fig. \ref{fig_SLGgap}(d). The presence of the Hubbard interaction enlarges the bulk band gap and weakens the finite size effect. Thus, the dispersion of the chiral edge states remains linear for a larger bandwidth, as shown by the spectral function in Fig. \ref{fig_SLGgap}(d).

\subsection{Topological phase diagram: effect of SOC}

The presence of SOC in single layer honeycomb lattice models with antiferromagnetic exchange field could induce topological phase transition or metal-insulator phase transition. Rotation of the antiferromagnetic exchange field between in-plane and out-of-plane direction could induce metal-insulator phase transition, so that the bulk could function as spin valve-I.

If the single layer hexagonal crystal, such as graphene, is in proximity to TMDCs, the Rashba SOC and staggered sublattice intrinsic SOC are induced \cite{Gmitra15,Gmitra16,Frank18}. The system is modeled by the Hamiltonian
\begin{equation}
H=H_{0}+H_{AF}+H_{R}+H_{sI}\label{HamiltonianSLGsI}
\end{equation}
The Rashba SOC is modeled by
\begin{equation}
H_{R}=\frac{2i\lambda_{R}}{3}\sum_{\langle i,j\rangle,\sigma,\sigma'}[(\hat{\mathbf{s}}\times\mathbf{d}_{ij})_{z}]_{\sigma,\sigma'}c_{i\sigma}^{+}c_{j\sigma'}
\label{HamiltonianR}
\end{equation}
where $\mathbf{d}_{ij}$ is the unit vector from the $i$-th to the $j$-th lattice site, and $\hat{\mathbf{s}}$ is the vector of three Pauli matrices for spin. The Rashba SOC can be tuned by the adatom doping of heavy metallic atoms \cite{Marchenko12}. The staggered sublattice intrinsic SOC is modeled by
\begin{equation}
H_{sI}=\frac{i\lambda_{sI}}{3\sqrt{3}}\sum_{\langle\langle i,j\rangle\rangle,\sigma,\sigma'}\kappa_{i}\nu_{ij}\hat{s}^{z}_{\sigma,\sigma'}c_{i\sigma}^{+}c_{j\sigma'}
\label{HamiltonianIs}
\end{equation}
where $\nu_{ij}=(+1)-1$ for a (counter)clockwise path. The summation indices with $\langle\langle i,j\rangle\rangle$ cover the next-nearest neighboring lattice sites. On the other hand, if the single layer hexagonal crystal is doped with adatom that induce Rashba SOC and uniform intrinsic SOC \cite{ConanWeeks12}, the system is modeled by the Hamiltonian
\begin{equation}
H=H_{0}+H_{AF}+H_{R}+H_{I}\label{HamiltonianSLGuI}
\end{equation}
The uniform intrinsic SOC is modeled by
\begin{equation}
H_{I}=\frac{i\lambda_{I}}{3\sqrt{3}}\sum_{\langle\langle i,j\rangle\rangle,\sigma,\sigma'}\nu_{ij}\hat{s}^{z}_{\sigma,\sigma'}c_{i\sigma}^{+}c_{j\sigma'}
\label{HamiltonianI}
\end{equation}
We separately study the phase diagrams of the two model Hamiltonians (\ref{HamiltonianSLGsI}) and (\ref{HamiltonianSLGuI}). Without the antiferromagnetic exchange field, the first Hamiltonian is in the quantum valley Hall (QVH) phase with $\mathcal{C}=0$ and $\mathcal{C}_{V}=2\cdot sign(\lambda_{sI})$; the second Hamiltonian is the well-known Kane-Mele-Rashba model, which is in either a trivial or the quantum spin Hall (QSH) phase \cite{Kane05}.

\begin{figure}[tbp]
\scalebox{0.58}{\includegraphics{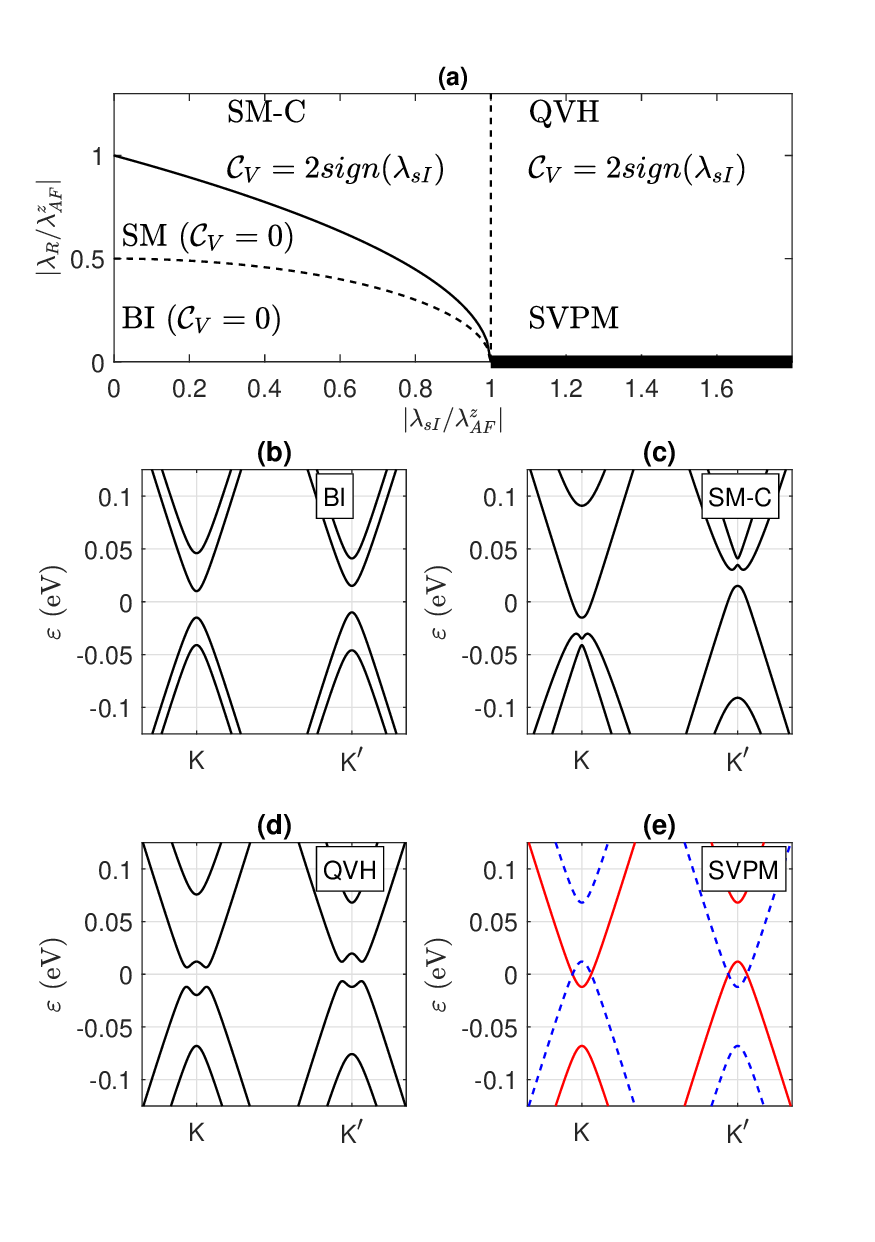}}
\caption{ (a) The phase diagram in the $\lambda_{R}-\lambda_{sI}$ space of the Hamiltonian $H_{0}+H_{AF}+H_{R}+H_{sI}$ with out-of-plane antiferromagnetic exchange field ($\lambda_{AF}^{x}=\lambda_{AF}^{y}=0,\lambda_{AF}^{z}=0.01t$). The typical band structures of band insulator (BI), semi-metal with nonzero Chern number (SM-C), quantum valley Hall (QVH) and spin-valley-polarized metal (SVPM) phases are plotted in (b), (c), (d) and (e), respectively. In (e), the band structures of the spin-up and spin-down components are plotted as red (solid) and blue (dashed) lines, respectively.  }
\label{fig_SLGphase1}
\end{figure}

\begin{figure}[tbp]
\scalebox{0.58}{\includegraphics{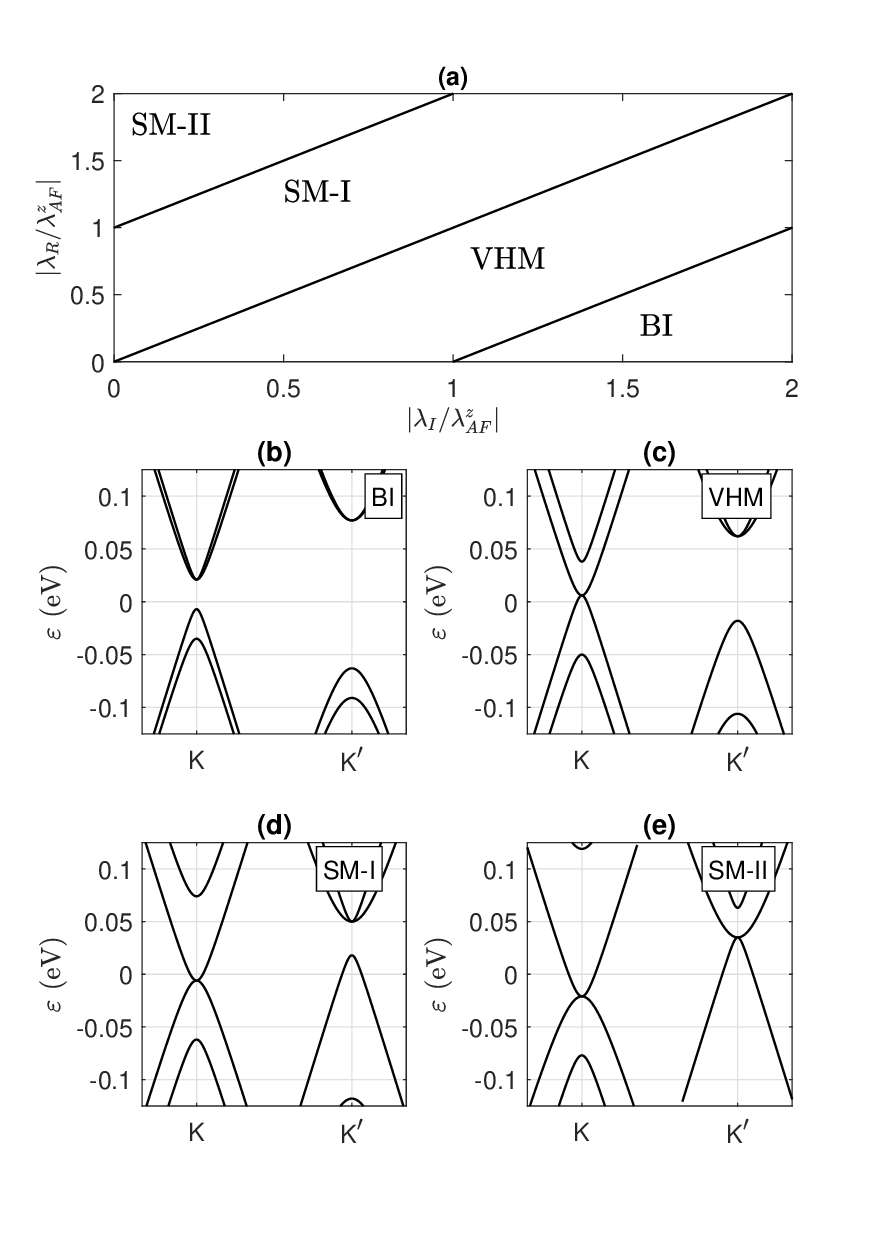}}
\caption{ (a) The phase diagram in the $\lambda_{R}-\lambda_{I}$ space of the Hamiltonian $H_{0}+H_{AF}+H_{R}+H_{I}$ with out-of-plane antiferromagnetic exchange field ($\lambda_{AF}^{x}=\lambda_{AF}^{y}=0,\lambda_{AF}^{z}=0.01t$). The typical band structures of each phase are plotted in (b) to (e). }
\label{fig_SLGphase2}
\end{figure}

The phase diagram of the model Hamiltonian (\ref{HamiltonianSLGsI}) with out-of-plane antiferromagnetic exchange field ($\lambda_{AF}^{x}=\lambda_{AF}^{y}=0,\lambda_{AF}^{z}=0.01t$) are plotted in Fig. \ref{fig_SLGphase1}(a). The solid curve $\lambda_{R}=\sqrt{\lambda_{AF}^{z}(\lambda_{AF}^{z}-\lambda_{sI})}$ separates the phase diagram into two regimes with $\mathcal{C}_{V}=2\cdot sign(\lambda_{sI})$ to the right and $\mathcal{C}_{V}=0$ to the left. Across the curve, the band gap at $K$ and $K^{\prime}$ closes and reopens. The regime with $\mathcal{C}_{V}=0$ is further separated into two regimes by the dashed curve $\lambda_{R}=\frac{1}{2}\sqrt{(\lambda_{AF}^{z})^{2}-\lambda_{sI}^{2}}$. To the left of the dashed curve, the global band gap is finite, and the systems are in the BI phase. The band structure of a typical system in the BI phase is plotted in Fig. \ref{fig_SLGphase1}(b). To the right of the dashed curve, the global band gap vanishes because the band gaps of the $K$ and $K^{\prime}$ valleys cover different energy ranges. The intrinsic Fermi level crosses the conductor(valence) band of the $K$($K^{\prime}$) valley, and this phase is thus designated the semi-medal (SM) phase. The regime with $\mathcal{C}_{V}=2\cdot sign(\lambda_{sI})$ is also further separated into two regimes by the vertical dashed line $\lambda_{sI}=\lambda_{AF}^{z}$. To the left of the dashed line, the global band gap vanishes. This phase regime is designated as SM with nonzero valley Chern number (SM-C). The band structure of a typical system in the SM-C phase is plotted in Fig. \ref{fig_SLGphase1}(c). To the right of the dashed line, the global band gap is finite, and the systems are in a QVH phase. The band structure of a typical system in the QVH phase is plotted in Fig. \ref{fig_SLGphase1}(d). When $\lambda_{R}$ is equal to zero in this regime (thick line in the phase diagram), a typical band structure is plotted in Fig. \ref{fig_SLGphase1}(e), which is gapless. The band structures of the two spin components have opposite valley polarization, and this phase is thus designated spin-valley-polarized metal (SVPM). If the absolute value of the Fermi level is tuned to be greater than $\lambda_{sI}-\lambda_{AF}^{z}$, the systems exhibit coupled spin and valley physics that is similar to that of the TMDCs \cite{DiXiao12}.

The phase diagram of the model Hamiltonian (\ref{HamiltonianSLGuI}) with out-of-plane antiferromagnetic exchange field ($\lambda_{AF}^{x}=\lambda_{AF}^{y}=0,\lambda_{AF}^{z}=0.01t$) is shown in Fig. \ref{fig_SLGphase2}(a). In the absence of $H_{AF}$, the Hamiltonian could be in QSH phase with topological number $\mathbb{Z}_{2}=1$ \cite{Kane05}. The presence of the antiferromagnetic exchange field breaks the time-reversal symmetry and drives the system into the topological trivial phase with $\mathcal{C}=0$ and $\mathcal{C}_{V}=0$. The four phases are characterized by global and valley band gaps. The typical band structure of each phase is plotted in Fig. \ref{fig_SLGphase2}(b-e). The three straight lines that separate the four phases are $\lambda_{R}=\lambda_{I}-\lambda_{AF}^{z}$, $\lambda_{R}=\lambda_{I}$ and $\lambda_{R}=\lambda_{I}+\lambda_{AF}^{z}$. The phase in Fig. \ref{fig_SLGphase2}(c) is valley half-metal (VHM), because the intrinsic Fermi level crosses the valence band of the $K$ valley. The phases in Fig. \ref{fig_SLGphase2}(d) and (e) are both SMs.

\begin{figure}[tbp]
\scalebox{0.43}{\includegraphics{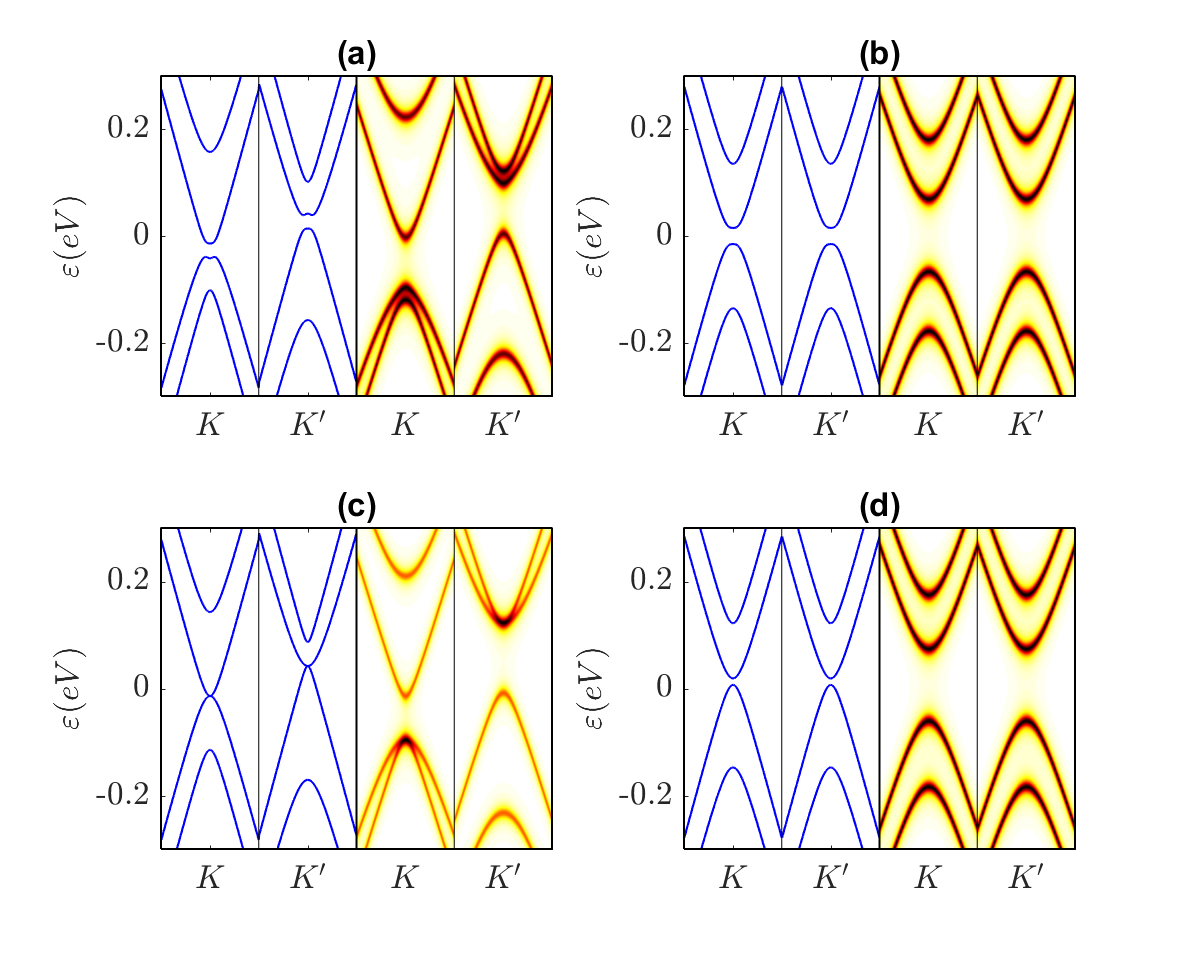}}
\caption{ The band structure and spectral function of single layer honeycomb lattice with out-of-plane ($\lambda_{AF}^{x}=\lambda_{AF}^{y}=0,\lambda_{AF}^{z}=0.01t$) and in-plane ($\lambda_{AF}^{y}=\lambda_{AF}^{z}=0,\lambda_{AF}^{x}=0.01t$) antiferromagnetic exchange field in the left and right columns, respectively. The strength of Rashba SOC is $\lambda_{R}=2.3\lambda_{AF}$ in all sub-figures. Staggered sublattice intrinsic SOC with $\lambda_{sI}=0.5\lambda_{AF}$ in (a) and (b); Uniform intrinsic SOC with $\lambda_{I}=0.5\lambda_{AF}$ in (c) and (d). }
\label{fig_SLGvalve}
\end{figure}

For the metallic phases (SMs and VHM) in Fig. \ref{fig_SLGphase1}(a) and \ref{fig_SLGphase2}(a), rotation of the antiferromagnetic exchange field to in-plane direction ($\lambda_{AF}^{y}=\lambda_{AF}^{z}=0,\lambda_{AF}^{x}=0.01t$) drives the systems into band insulator, whose band gaps of the two valleys have the same energy range. Thus, spin valve-I could be constructed by utilizing the metal-insulator transition of the systems in these phases. In the addition presence of Hubbard interaction, the band gap of each valley is either opened or enlarged. Consequently, if the strength of the Rashba SOC is small ($\lambda_{R}<2.3\lambda_{AF}$), the metallic phases in Fig. \ref{fig_SLGphase1}(a) and \ref{fig_SLGphase2}(a) become insulator, so that the antiferromagnetic driven metal-insulator transition is absent. The band structures of noninteracting models and the spectral functions of the interacting models of two typical systems with Rashba SOC strength being large enough ($\lambda_{R}=2.3\lambda_{AF}$) are shown in Fig. \ref{fig_SLGvalve}. The two systems function as spin valve-I. For the system with Hamiltonian (\ref{HamiltonianSLGsI}) and out-of-plane antiferromagnetic exchange field, the spectral function in Fig. \ref{fig_SLGvalve}(a) shows that the band gap of each valley is enlarged by Hubbard interaction, while the global band gap remain being zero. For the same system with in-plane antiferromagnetic exchange field, Hubbard interaction enlarges the global band gaps, as shown in Fig. \ref{fig_SLGvalve} (b). Thus, the sensitivity of the band gap to the rotation of the antiferromagnetic exchange field is enhanced by Hubbard interaction. Similar phenomenons are found for the systems with Hamiltonian (\ref{HamiltonianSLGuI}) in Fig. \ref{fig_SLGvalve} (c) and (d). In Fig. \ref{fig_SLGvalve} (c), the band gap in each valley is opened by Hubbard interaction.

\section{bilayer honeycomb lattice}

For bilayer honeycomb lattice, the exchange fields could have three configurations: (i) antiferromagnetic heterostructure with antiferromagnetic exchange fields for both layers; (ii) antiferromagnetic/ferromagnetic heterostructure with ferromagnetic exchange field for top layer and antiferromagnetic exchange field for bottom layer; (iii) ferromagnetic heterostructure with ferromagnetic exchange fields for both layers. The Hamiltonian for the three configurations are
\begin{equation}
H=\sum_{\iota=1,2}{(H_{0,\iota}+H_{AF,\iota})}+H_{\bot}\label{HamiltonianBLG_AA}
\end{equation}
\begin{equation}
H=\sum_{\iota=1,2}{(H_{0,\iota})}+H_{FM,1}^{z}+H_{AF,2}+H_{\bot}\label{HamiltonianBLG_AF}
\end{equation}
\begin{equation}
H=\sum_{\iota=1,2}{(H_{0,\iota}+H_{FM,\iota}^{z})}+H_{\bot}\label{HamiltonianBLG_FF}
\end{equation}
where $\iota$ is an additional layer index with $\iota=1(2)$ standing for top(bottom) layer. The interlayer hopping is described by the Hamiltonian
\begin{equation}
H_{\bot}=t_{\bot}\sum_{\langle i,j\rangle_{\bot},\sigma}c_{i\sigma}^{+}c_{j\sigma}
\end{equation}
where $t_{\bot}=0.39$ eV, the summation indices $\langle i,j\rangle_{\bot}$ couple the B lattice sites of the $\iota=1$ layer and the nearest A lattice site of the $\iota=2$ layer. The ferromagnetic exchange field, given as
\begin{equation}
H_{FM}^{x,y,z}=\lambda_{FM}^{z}\sum_{i,\sigma,\sigma'}\hat{s}^{z}_{\sigma,\sigma'}c_{i\sigma}^{+}c_{i\sigma'}
\label{HamiltonianFM}
\end{equation}
, could be induced by proximity to ferromagnetic substrate. The coefficients $\lambda_{FM}^{z}$ could be extracted from the density functional theory (DFT) calculation \cite{Cardoso18,HXYang13,Hallal17,Zollner18}.

The ferromagnetic heterostructure is conductor (insulator) if the exchange fields of the two graphene layers are parallel (antiparallel) \cite{Cardoso18}. The antiferromagnetic and antiferromagnetic/ferromagnetic heterostructures exhibit a similar property to function as spin valve-I: the bulk band gap depends on the magnetization orientation of the antiferromagnetic exchange field. The following four subsections discuss the details of four issues: (i) the presence of Hubbard interaction changes the band gap or induces metal-insulator phase transition; (ii) the presence of Rashba SOC drives the topological phase transition to Chern insulator; (iii) the coexistence of Hubbard interaction and Rashba SOC enlarge the band gap of the Chern insulator; and (iv) the domain wall between two heterostructures with different Chern numbers supports topological edge states.

\subsection{Bulk band gap: effect of Hubbard interaction}

\begin{figure}[tbp]
\scalebox{0.36}{\includegraphics{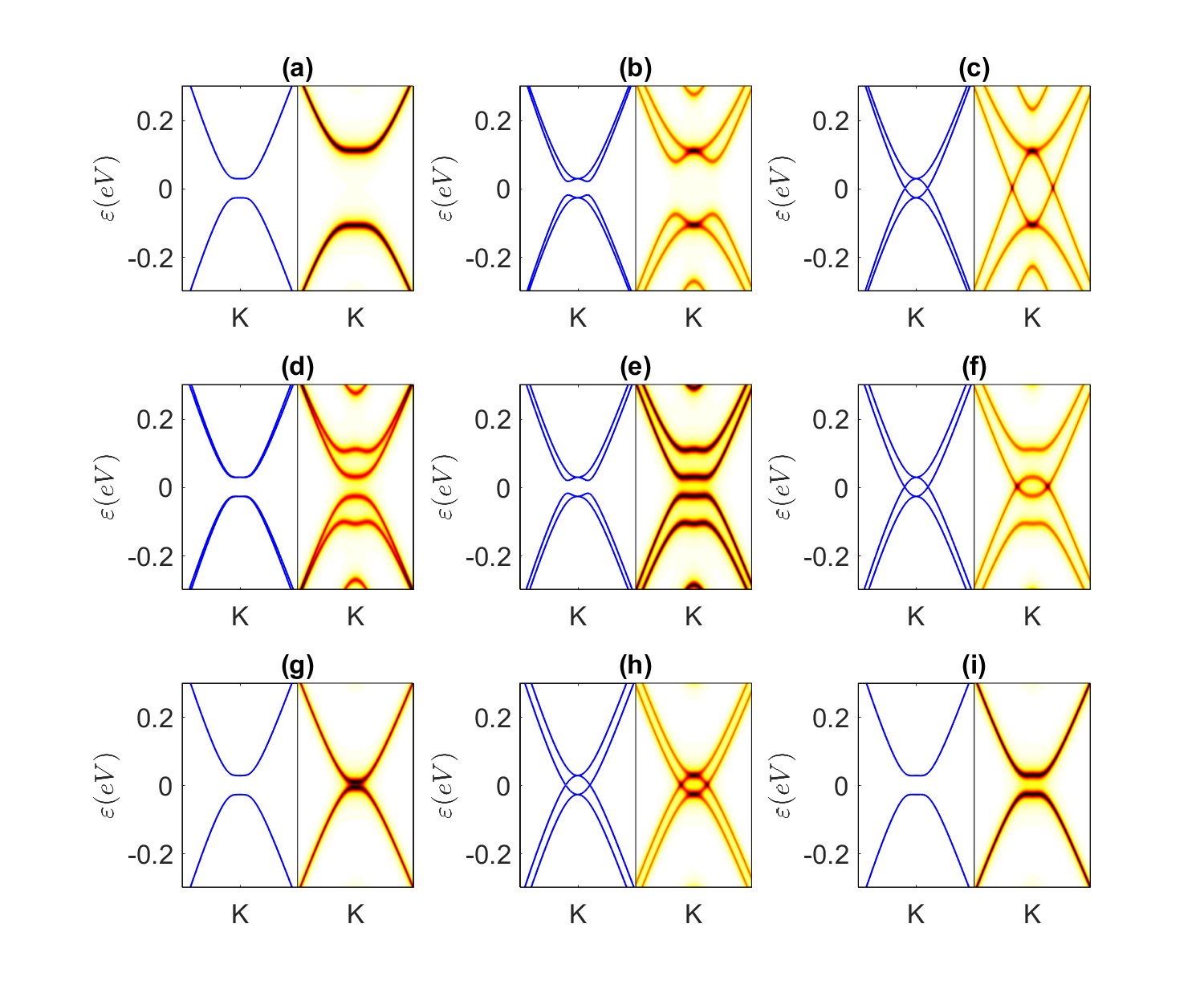}}
\caption{ The band structure and spectral function of the antiferromagnetic heterostructures, with $\lambda_{AF,1}^{z}=\lambda_{AF,2}^{z}=0.01t$ in (a), $\lambda_{AF,1}^{z}=\lambda_{AF,2}^{x}=0.01t$ in (b) and $\lambda_{AF,1}^{z}=-\lambda_{AF,2}^{z}=0.01t$ in (c). The band structure and spectral function of the antiferromagnetic/ferromagnetic heterostructures, with $\lambda_{FM,1}^{z}=\lambda_{AF,2}^{z}=0.01t$ in (d), $\lambda_{FM,1}^{z}=\lambda_{AF,2}^{x}=0.01t$ in (e) and $\lambda_{FM,1}^{z}=-\lambda_{AF,2}^{z}=0.01t$ in (f). The band structure and spectral function of bilayer honeycomb lattice with nonmagnetic staggered sublattice potential with $\lambda_{\Delta,1}=\lambda_{\Delta,2}=0.01t$ in (g). The band structure and spectral function of the ferromagnetic heterostructures, with $\lambda_{FM,1}^{z}=\lambda_{FM,2}^{z}=0.01t$ in (h) and $\lambda_{FM,1}^{z}=-\lambda_{FM,2}^{z}=0.01t$ in (i). For all heterostructures in this figure, the SOCs are absent.  }
\label{fig_BLGgap}
\end{figure}

The presence of Hubbard interaction is modeled by the additional Hubbard terms in the Hamiltonian (\ref{HamiltonianBLG_AA}), (\ref{HamiltonianBLG_AF}) and (\ref{HamiltonianBLG_FF}), whose spectral functions are calculated by CPT. For varying types of hererostructures, the band structures of noninteracting models and the spectral functions of the interacting models are plotted in Fig. \ref{fig_BLGgap}.

For the antiferromagnetic heterostructures, the exchange fields of the two graphene layers are parallel, perpendicular and antiparallel to each other in Fig. \ref{fig_BLGgap}(a), (b) and (c), respectively. The heterostructures were switched from insulator to conductor by rotating the magnetization orientation of one of the two antiferromagnetic substrates. For the insulator phase, as shown in Fig. \ref{fig_BLGgap}(a) and (b), the Hubbard interaction enlarges the band gap, because the self-energy effectively enhance the spin-dependent staggered sublattice potential.

\begin{figure}[tbp]
\scalebox{0.56}{\includegraphics{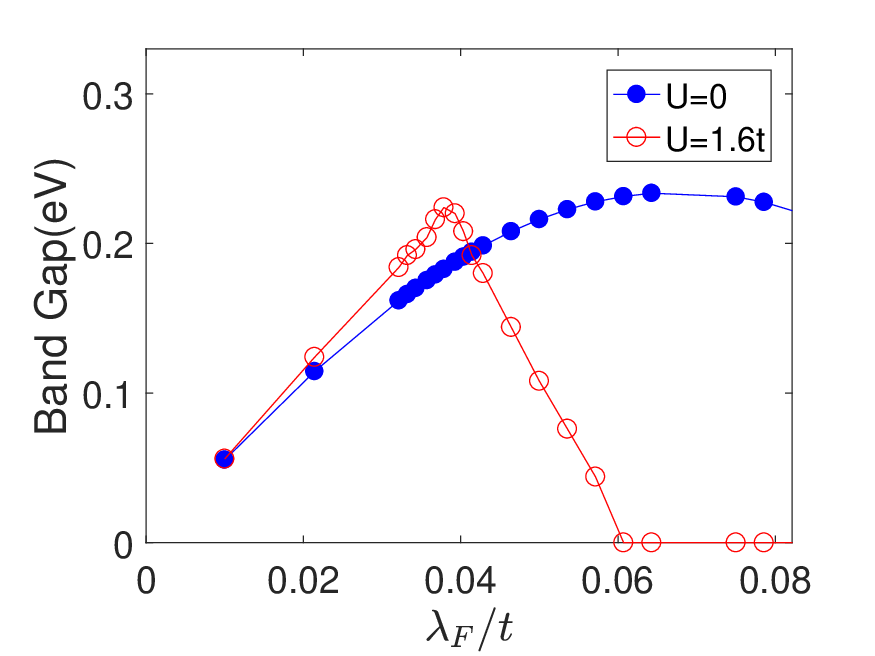}}
\caption{ For antiferromagnetic/ferromagnetic heterostructures, the bulk band gap versus $\lambda_{FM,1}^{z}=\lambda_{AF,2}^{z}=\lambda_{F}$ with and without Hubbard interaction is plotted as open and filled circles. }
\label{fig_BLGgapsize}
\end{figure}

For the antiferromagnetic/ferromagnetic heterostructures, the band gap is tuned by rotating the magnetization orientation of the antiferromagnetic exchange field while fixing the ferromagnetic exchange field, as shown in Fig. \ref{fig_BLGgap}(d-f). For the insulator phase, the Hubbard interaction slightly enlarges the band gap (Fig. \ref{fig_BLGgap}(d)). For the heterostructures with larger exchange fields, i.e., larger values of $\lambda_{F}$($=\lambda_{FM,1}^{z}=\lambda_{AF,2}^{z}$), the Hubbard interaction more significantly enlarges the band gap, as shown in Fig. \ref{fig_BLGgapsize}. However, with $\lambda_{F}$ being larger than $0.036t$, the trend reverses, and the Hubbard interaction significantly suppresses the band gap. When the strength of $\lambda_{F}$ reaches the critical value at $0.061t$, the band gap is closed by the Hubbard interaction, implying that the insulator to metal phase transition is driven by the Hubbard interaction.

For comparison, the band structure of bilayer honeycomb lattice with nonmagnetic staggered sublattice potential is plotted in Fig. \ref{fig_BLGgap}(g), which shows that the Hubbard interaction suppresses the band gap \cite{JinRongXu161}. For ferromagnetic heterostructures, the Hubbard interaction does not significantly change the band gap, as shown in Fig. \ref{fig_BLGgap}(i). In conclusion, with the same coupling strength, the Hubbard interaction makes antiferromagnetic heterostructures more sensitive to the rotation of magnetization orientation than ferromagnetic heterostructures.

\subsection{Topological phase diagram: effect of SOC}

\begin{figure}[tbp]
\scalebox{0.36}{\includegraphics{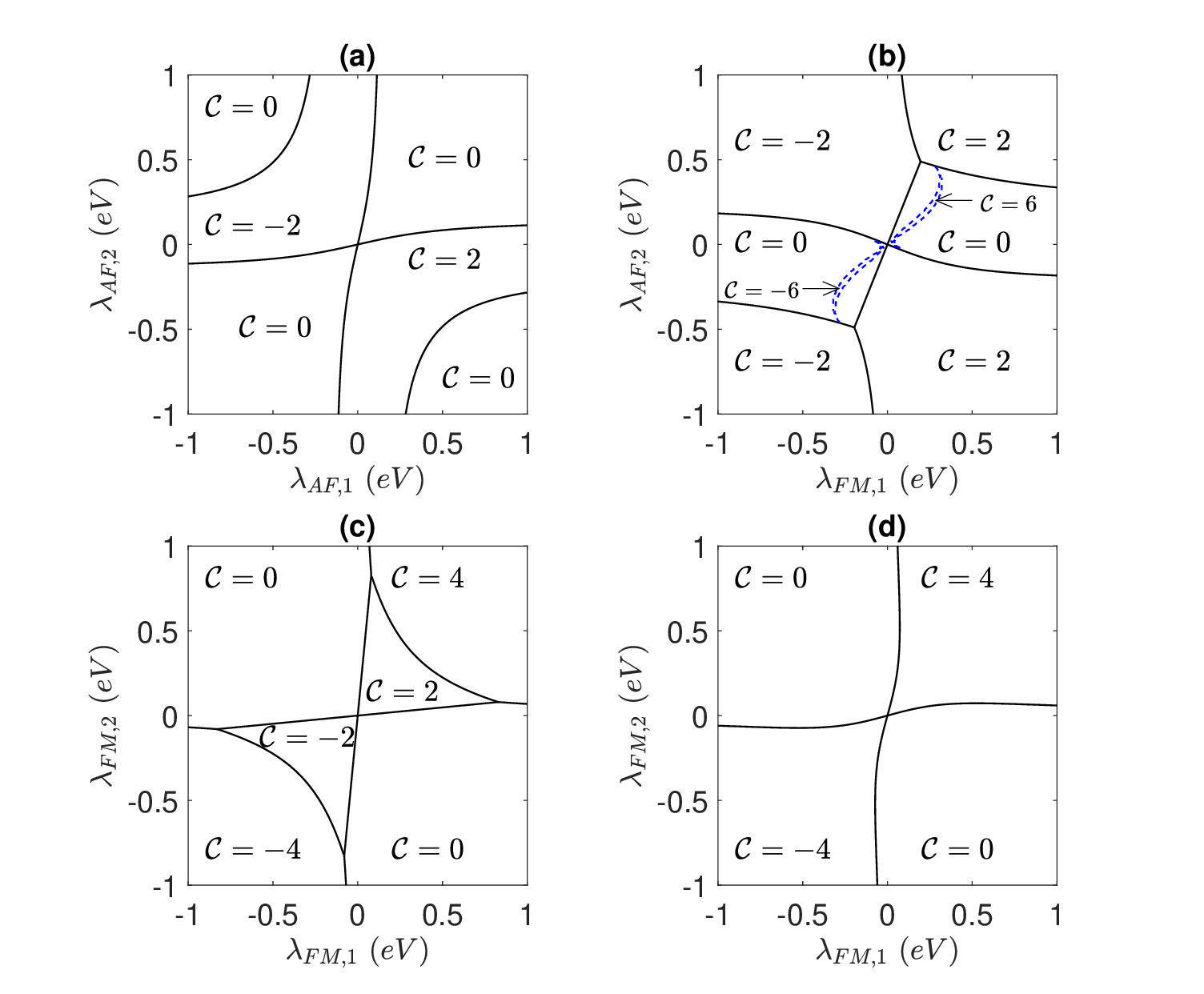}}
\caption{ The phase diagram of the Chern number of (a) antiferromagnetic, (b) antiferromagnetic/ferromagnetic and (c-d) ferromagnetic heterostructures. The Rashba SOC strength is $\frac{2}{3}\lambda_{R}=0.1$ eV in (a), $\frac{2}{3}\lambda_{R}=0.15$ eV in (b), $\frac{2}{3}\lambda_{R}=0.06$ eV $<\frac{1}{3}t_{\bot}$ in (c), and $\frac{2}{3}\lambda_{R}=0.15$ eV $>\frac{1}{3}t_{\bot}$ in (d). }
\label{fig_BLGphase}
\end{figure}

The presence of SOC in the bilayer honeycomb lattice models modifies the topological phase. Because both surfaces of the bilayer hexagonal crystals are in proximity to (anti)ferromagnetic insulators, no proximity to TMDCs is feasible. Thus, we neglect the presence of intrinsic SOC. Rashba SOC could be induced by the intercalation of heavy metallic atoms between two graphene layers \cite{Marchenko12}. We assumed that the Rashba SOC strengths of the two hexagonal crystal layers are the same, i.e., $H_{R,1}=H_{R,2}$. The phase diagrams of antiferromagnetic, antiferromagnetic/ferromagnetic and ferromagnetic heterostructures with Rashba SOC, which are modeled by the Hamiltonian
\begin{equation}
\sum_{\iota}{(H_{0,\iota}+H_{AF,\iota}^{z}+H_{R,\iota})}+H_{\bot}
\end{equation}
\begin{equation} \sum_{\iota}{(H_{0,\iota}+H_{R,\iota})}+H_{FM,1}^{z}+H_{AF,2}^{z}+H_{\bot}
\end{equation}
and
\begin{equation}
\sum_{\iota}{(H_{0,\iota}+H_{FM,\iota}^{z}+H_{R,\iota})}+H_{\bot}
\end{equation}
are plotted in Fig. \ref{fig_BLGphase}(a), (b) and (c-d), respectively.

For the antiferromagnetic heterostructures, the phase boundary near the x axis is obtained by solving the Hamiltonian at point $K$, which is given as
\begin{equation}
\tilde{\lambda}_{AF,2}^{z}=\frac{2\tilde{\lambda}_{AF,1}^{z}-2(\tilde{\lambda}_{AF,2}^{z})^{3}+\tilde{\lambda}_{AF,1}^{z}(\tilde{t}_{\perp})^{2}}{2[1-(\tilde{\lambda}_{AF,2}^{z})^{2}]}
\end{equation}
where $\tilde{\lambda}_{AF,1(2)}^{z}=\lambda_{AF,1(2)}^{z}/\lambda_{R}$, $\tilde{t}_{\perp}=t_{\perp}/\lambda_{R}$. Along this phase boundary, the band crossing at $K$ point has energy $\lambda_{AF,1}^{z}$. The phase boundaries near the y axis have similar features. The phase boundaries around the corner of the phase diagram are numerically calculated. For the antiferromagnetic/ferromagnetic and ferromagnetic heterostructures, there are two types of phase boundaries. Along the phase boundaries with solid(black) lines in Fig. \ref{fig_BLGphase}(b), the band crossing occurs at point $K$. The phase boundaries are either analytically expressed or numerically calculated. Along the phase boundaries with dashed (blue) lines, the band crossing occurs at three points that are at the three high symmetric $K-M$ lines with the same distance from the $K$ point \cite{maluo18blg}. The phase regimes that are surrounded by this type of phase boundaries have a Chern number of $\mathcal{C}=\pm6$. The shape of these phase regimes depends strongly on the strength of the Rashba SOC. For the ferromagnetic heterostructures, the phase diagrams for the systems with $\lambda_{R}<t_{\perp}/2$ and $\lambda_{R}>t_{\perp}/2$ are different, as shown in Fig. \ref{fig_BLGphase}(c) and (d), respectively. Both phase diagrams contain regimes with a large Chern number, $\mathcal{C}=\pm4$. The former phase diagram contains additional regimes with $\mathcal{C}=\pm2$ near the origin. For the regimes with nonzero Chern number and a finite global band gap, the heterostructures are in the QAH phase with quantized charge Hall conductance of $\sigma_{yx}=\mathcal{C}e^{2}/h$ \cite{ZhenhuaQiao10,MotohikoEzawa12}. For the antiferromagnetic and antiferromagnetic/ferromagnetic heterostructures, flipping the antiferromagnetic exchange field of the bottom layer ($\lambda_{AF,2}^{z}\Leftrightarrow-\lambda_{AF,2}^{z}$) could switch the systems between QAH and BI phases. The nanoribbon of such systems are switched between having and not having robust edge states by flipping the antiferromagnetic exchange field, which function as spin valve-II.

\subsection{Coexistence of Hubbard interaction and Rashba SOC}

Although the phase diagrams of the noninteracting model in Fig. \ref{fig_BLGphase} contain regimes with nonzero Chern number, some systems in these phase regimes have small or vanishing global band gap because the band gaps of the two valleys have different energy ranges. If the global band gap vanishes, the system is not in the QAH phase but is designated as a Chern topological metallic (CTM) phase. In the phase diagrams (Fig. \ref{fig_BLGphase}), the CTM phase would distribute within the regimes with nonzero Chern number and with small magnitude of the exchange fields. For the noninteracting model of the CTM phase, the band gap of each valley is small, and thus, the topological edge states at the open boundary \cite{HuiPan14} are weakly localized.

\begin{figure}[tbp]
\scalebox{0.53}{\includegraphics{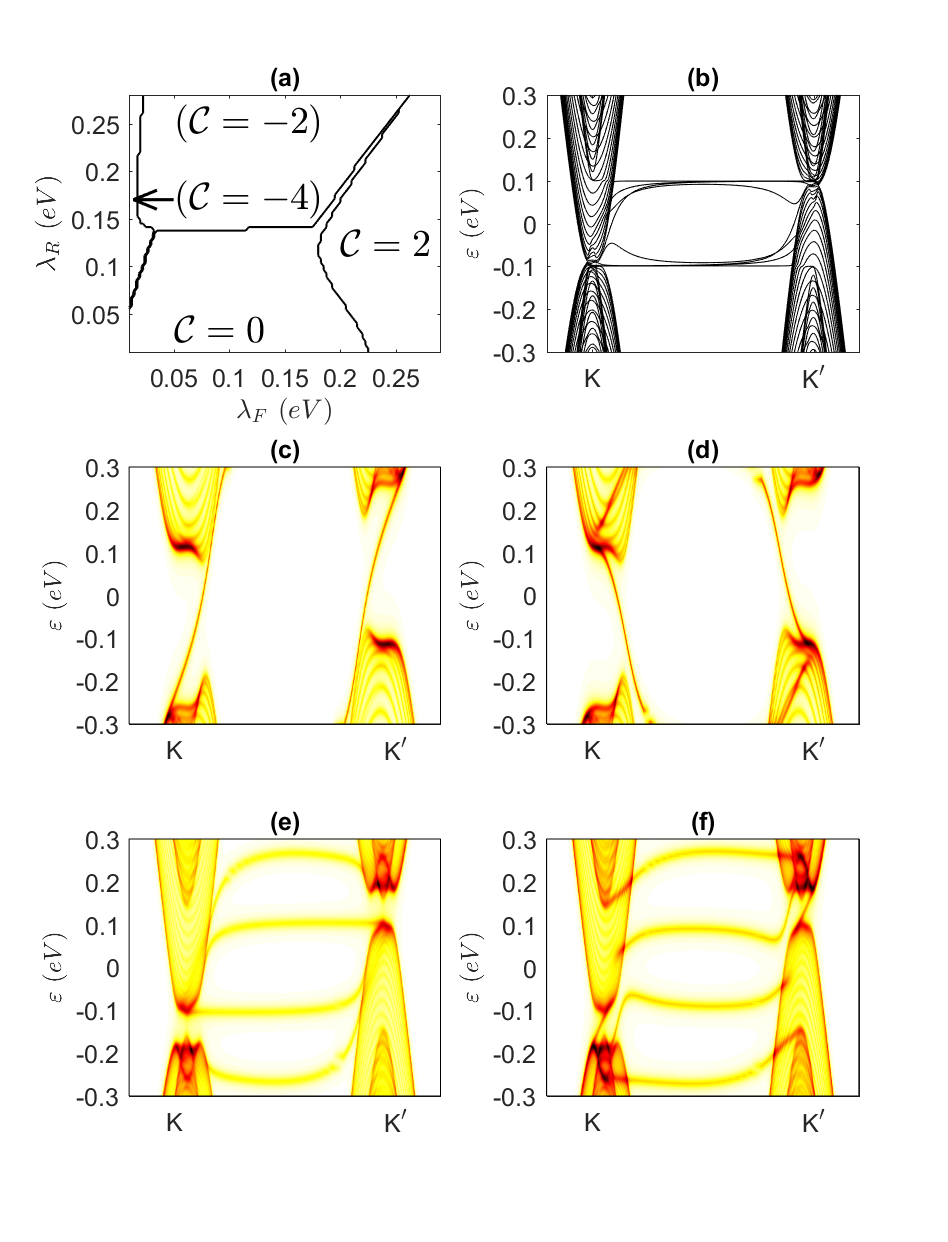}}
\caption{ (a) The phase diagram of the interacting model of antiferromagnetic/ferromagnetic heterostructures with $\lambda_{FM,1}^{z}=\lambda_{AF,2}^{z}=\lambda_{F}$ versus  $\lambda_{R}$. For the zigzag nanoribbon of the antiferromagnetic/ferromagnetic heterostructures with $\lambda_{FM,1}^{z}=\lambda_{AF,2}^{z}=0.4$ eV and $\lambda_{R}=0.2$ eV, (b) is the band structure of the noninteracting model, and (c) and (d) are the spectral functions of the interacting model of the left and right edges, respectively. The total width of the ribbon is $21.3$ nm. For another nanoribbon with $\lambda_{FM,1}^{z}=\lambda_{AF,2}^{z}=0.1$ eV and $\lambda_{R}=0.2$ eV, (e) and (f) are the spectral functions of the left and right edges, respectively. The total width of the ribbon is $42.6$ nm. }
\label{fig_BLGchernRibbon}
\end{figure}

In the additional presence of Hubbard interaction, the phase boundaries are modified; the global band gap and the band gap of each valley are enlarged. We use the antiferromagnetic/ferromagnetic heterostructures with $\lambda_{FM,1}^{z}=\lambda_{AF,2}^{z}=\lambda_{F}$ as an example to demonstrate these features. The Chern number is calculated by integrating the Berry curvature of the topological Hamiltonian \cite{ZhongWang2010,Gurarie2011,ZhongWang12,ZhongWang122,ZhongWang13,YuanYao16}, which is the inverse of the Green's function at zero frequency given by the CPT method. The numerical result of the phase diagram is plotted in Fig. \ref{fig_BLGchernRibbon}(a). Three regimes with Chern numbers of $+2$, $-2$ and $-4$ are identified. In the regimes with $\mathcal{C}=-2$ or $\mathcal{C}=-4$, the global band gaps vanish, leaving the systems in CTM phase. In the regime with $\mathcal{C}=2$, the global band gaps are finite.

To illustrate the topological global band gap that is enlarged by the Hubbard interaction, the zigzag nanoribbon of the heterostructures with $\lambda_{FM,1}^{z}=\lambda_{AF,2}^{z}=0.4$ eV and $\lambda_{R}=0.2$ eV is numerically investigated. The noninteracting model of this heterostructures is trivial (i.e., $\mathcal{C}=0$). The band structure of the nanoribbon is plotted in Fig. \ref{fig_BLGchernRibbon}(b), which exhibits features of SM. The interacting model of the same heterostructures is nontrivial (i.e., $\mathcal{C}=2$). The spectral functions of two edges that are obtained by spatially integrating the localized spectral function over the left and right halves of the nanoribbon, are plotted in Fig. \ref{fig_BLGchernRibbon}(c) and (d), respectively. Two chiral edge states appear at each edge with energy within the global bulk band gap.

The presence of the Hubbard interaction also enlarges the band gap of each valley in the CTM phase. For the heterostructures with $\lambda_{FM,1}^{z}=\lambda_{AF,2}^{z}=0.1$ eV and $\lambda_{R}=0.2$ eV, the interacting model is nontrivial with $\mathcal{C}=-2$. The spectral functions of the left and right edges of the zigzag nanoribbon of the heterostructures are plotted in Fig. \ref{fig_BLGchernRibbon}(e) and (f), respectively. Within the band gap of each valley, two edge states localized at the right open boundary appear, as shown in Fig. \ref{fig_BLGchernRibbon}(f). The edge states that connects the conductor and valence bulk bands in the same valley are topologically protected. Although the global band gap vanishes, the novel transportation feature of the topologically protected edge states could be harnessed by tuning the Fermi level into the band gap of one valley.

\subsection{Topological edge states at domain wall}

\begin{figure}[tbp]
\scalebox{0.58}{\includegraphics{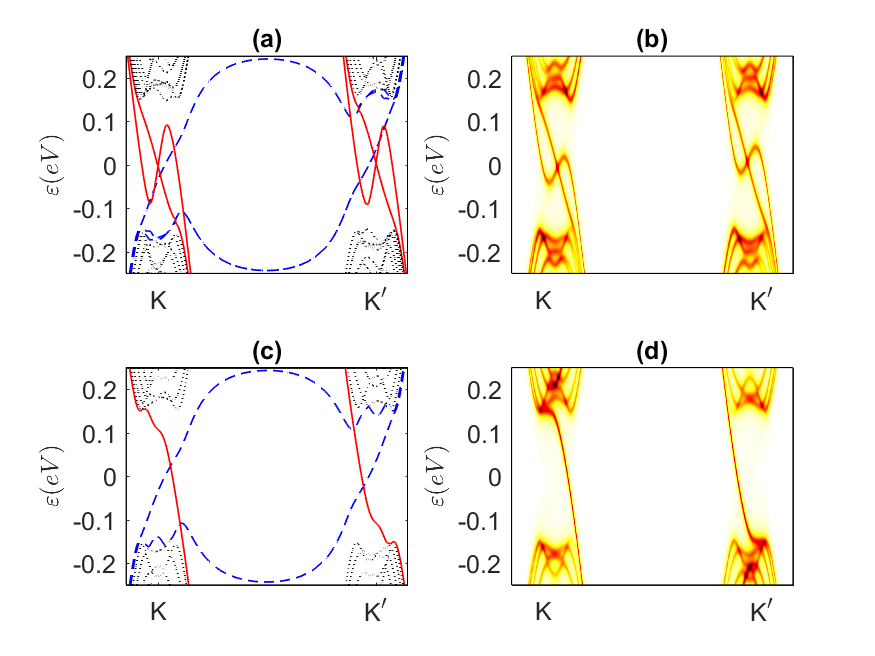}}
\caption{ Band structure (left column) and spectral function (right column) of the zigzag nanoribbon of antiferromagnetic heterostructures with domain wall in the middle. The total width of the nanoribbon is $42.6$ nm. The left half of the nanoribbon has parameters $\lambda_{AF,1}^{z}=-\lambda_{AF,2}^{z}=0.28$ eV, $2\lambda_{R}/3=0.3$ eV and $\mathcal{C}=2$. In (a-b), the right half of the nanoribbon has parameters $\lambda_{AF,1}^{z}=-\lambda_{AF,2}^{z}=-0.28$ eV, $2\lambda_{R}/3=0.3$ eV and $\mathcal{C}=-2$. In (c-d), the right half of the nanoribbon has parameters $\lambda_{AF,1}^{z}=\lambda_{AF,2}^{z}=0.28$ eV, $2\lambda_{R}/3=0$ eV and $\mathcal{C}=0$. For the band structure of the noninteracting model, the chiral edge states at the domain wall are plotted as solid (red) lines; the edge states at the open boundaries are plotted as dashed (blue) lines; the bulk states are plotted as dotted (black) lines. }
\label{fig_BLGribbon1}
\end{figure}

The domain walls between two regions with different Chern numbers and overlapping bulk gaps host localized chiral edge states. Nanoribbons with such domain wall along the axis function as spin valve-III. The first (second) type of domain wall has Chern numbers of $\mathcal{C}=\pm2$ ($\mathcal{C}=2$ and $\mathcal{C}=0$) on either side. For a typical zigzag nanoribbon of antiferromagnetic heterostructures with the first type of domain wall in the middle, the band structure of the noninteracting model and the spectral function of the interacting model are plotted in Fig. \ref{fig_BLGribbon1} (a) and (b), respectively. The exchange fields of both graphene layers are reversed across the domain wall. For the interacting model, the spectral function is obtained by spatially integrating the localized spectral function. The integral region avoids the open boundaries and covers the domain wall, so that only the chiral edge states of the domain wall are visualized. Because the Chern number is changed by four across the domain wall, there are four chiral edge states. Two of these chiral edge states have nearly linear monotonic dispersion; the other two chiral edge states have oscillating dispersion around the intrinsic Fermi level. In the presence of the Hubbard interaction, the band width of the oscillating dispersion is suppressed, as shown in Fig. \ref{fig_BLGribbon1}(b). For a typical zigzag nanoribbon of antiferromagnetic heterostructure with the second type of domain wall in the middle, the band structure and spectral function are plotted in Fig. \ref{fig_BLGribbon1} (c) and (d), respectively. The exchange field of only one graphene layer is reversed across the domain wall. The Rashba SOC of the region with $\mathcal{C}=0$ is set to zero, so that the bulk gap is large. Only two chiral edge states with nearly linear monotonic dispersion appear at the domain wall. For realistic materials, the parameters are smaller than those in Fig. \ref{fig_BLGribbon1}, so that the finite size effect couples the chiral edge states at the open boundaries and the domain wall. However, the coupling becomes insignificant for wider nanoribbons.

\begin{figure}[tbp]
\scalebox{0.58}{\includegraphics{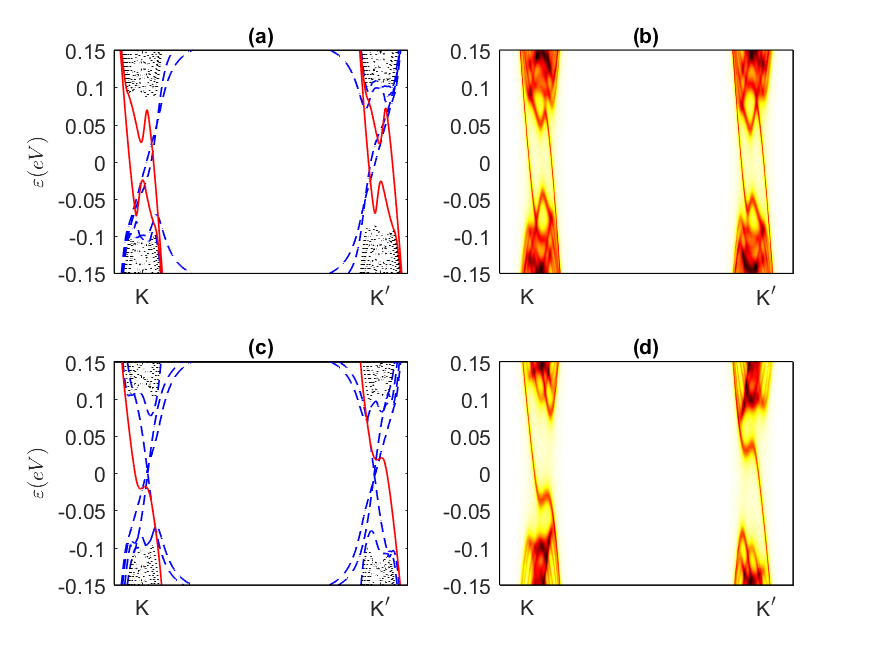}}
\caption{ The same type of plot as in Fig, \ref{fig_BLGribbon1}. The total width of the nanoribbon is $42.6$ nm. The left half of the nanoribbon has parameters $\lambda_{FM,1}^{z}=-\lambda_{AF,2}^{z}=0.17$ eV, $2\lambda_{R}/3=0.082$ eV and $\mathcal{C}=2$. In (a-b), the right half of the nanoribbon has parameters $\lambda_{FM,1}^{z}=-\lambda_{AF,2}^{z}=-0.17$ eV, $2\lambda_{R}/3=0.082$ eV and $\mathcal{C}=-2$. In (c-d), the right half of the nanoribbon has parameters $\lambda_{FM,1}^{z}=\lambda_{AF,2}^{z}=0.17$ eV, $2\lambda_{R}/3=0$ eV and $\mathcal{C}=0$. }
\label{fig_BLGribbon2}
\end{figure}

For a zigzag nanoribbon of antiferromagnetic/ferromagnetic heterostructures with the first and second types of domain wall in the middle, the numerical results are plotted in Fig. \ref{fig_BLGribbon2}(a-b) and (c-d), respectively. For the first type of domain wall, the dispersions of the four chiral edge states are nearly linear and monotonic within a bandwidth around the intrinsic Fermi level and are oscillating dispersively outside of this band width. The presence of the Hubbard interaction enlarges the bandwidth in which the chiral edge states have nearly linear monotonic dispersion, as shown in Fig. \ref{fig_BLGribbon2}(b). The two chiral edge states of the second type of domain wall have similar features.

\begin{figure}[tbp]
\scalebox{0.58}{\includegraphics{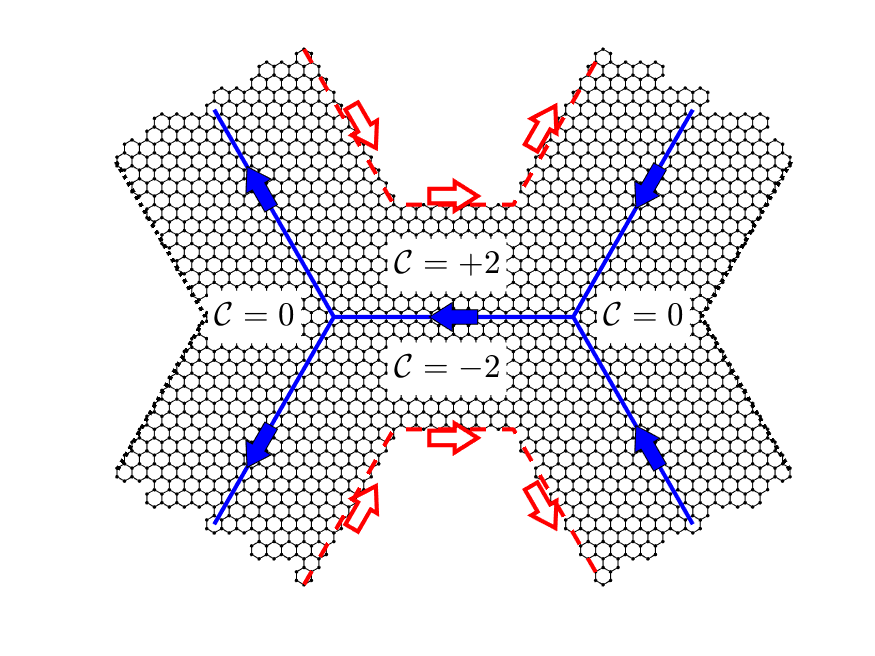}}
\caption{ Scheme of a three-way current partition and recombination at the junction of three domain walls. }
\label{fig_BLGsplit}
\end{figure}

Because the Chern number could be one of the three values, $\pm2$ or $0$, the junction among three regions with three different Chern numbers can be constructed, as shown in Fig. \ref{fig_BLGsplit}. The Chern number of each region is labeled in the figure. The junction of three domain walls supports current partition or recombination \cite{maluo3w1,maluo3w2}. In Fig. \ref{fig_BLGsplit}, the topological domain walls are marked by solid (blue) lines, with a solid (blue) arrow indicating the direction of the one-way charge current at Fermi level. The three-way junction at the left supports current partition, and that at the right supports current recombination. The open boundaries of the regions with $\mathcal{C}=\pm2$ are marked by red (dashed) lines, which also support one-way charge current labeled by the empty red arrows. The open boundaries of the region with $\mathcal{C}=0$ are marked by black (dotted) lines. On the other hand, the junction among four regions with Chern numbers of $+2$, $-2$, $+2$ and $-2$ along a (counter)clockwise sequence support the current partition and recombination as well \cite{Yafei17}. The large-scale integrated spintronic circuits could be constructed from jigsaws of regions with varying Chern numbers.

\section{conclusion}

Spin valves with CIP geometry consisting of heterostructures of two-dimensional hexagonal crystals and (anti)ferromagnetic insulators are studied by solving the tight binding model with the presence of a (staggered sublattice) exchange field. The band structure of heterostructures in bulk can be switched between insulating and metallic by rotating the magnetization orientation of the antiferromagnetic exchange field(s). The presence of Hubbard interaction increases the sensitivity of the band gap to the antiferromagnetic exchange field(s). In the presence of Rashba SOCs in the single layer honeycomb lattice model with out-of-plane antiferromagnetic exchange field, topological phase transitions among BI, SM-C, QVH and SVPM phases are driven by the staggered sublattice intrinsic SOC, while metal-insulator transition among BI, VHM and SM phases are driven by the uniform intrinsic SOC. Rotating the antiferromagnetic exchange field to in-plane direction switches the metallic phases into insulating phase. For antiferromagnetic, antiferromagnetic/ferromagnetic and ferromagnetic heterostructures of bilayer honeycomb lattice model, the Rashba SOC induces a topological phase transition to QAH or CTM phase with a Chern number of $\pm2$, $\pm4$ or $\pm6$. The coexistence of Hubbard interaction and Rashba SOC enlarges the global band gap of the QAH phase or the band gap of each valley of the CTM phase. According to the phase diagrams of the varying types of bilayer honeycomb lattice models, flipping the antiferromagnetic exchange field could switch the topological number between nonzero and zero, which in turn switch the topological edge states of nanoribbon on and off. Domain walls between two regions with different Chern numbers (or valley Chern numbers) and an overlapping energy range of the global band gaps host chiral edge states. The presence of Hubbard interaction improves the quality of varying types of chiral edge states by enlarging the band width with nearly linear monotonic dispersion. Three-way current partition or recombination devices consisting of the junction of three regions with three different Chern numbers (or the intersection of three domain walls) are proposed, which could be the building block for large-scale integrated antiferromagnetic spintronic systems. Further studies of antiferromagnetic spin valve systems would require the discovery or engineering of realistic antiferromagnetic materials to induce antiferromagnetic exchange fields, as well as additional adatom doping to induce SOC.

\section{Appendix}

Implementation of CPT follows the four steps to calculate the self-energy $\Sigma(\mathbf{k},z)$ in Eq. (\ref{greenfunction}):

(i) At first, a cluster that contain finite number of lattice sites is defined. The cluster must be able to tile the extended honeycomb lattice of the original model by being arranged in superlattice. Usually, the superlattice breaks the translation symmetry of the original honeycomb lattice.

(ii) Secondly, the Green's function of the isolated cluster is calculated by exact diagonalization (ED) of the Hamiltonian, $\mathcal{H}$, of the isolated cluster. The Hamiltonian includes the Hubbard model and the noninteracting model except for the hopping terms between different cluster. The matrix form of $\mathcal{H}$ is defined in the basis of all Fock states in the Hilbert space. For half filling of an N-site lattice with spin-$\frac{1}{2}$ Fermions, the number of Fock states equates to the combination $C_{N}^{2N}$. The ground state of the Hamiltonian, $|\Omega\rangle$, could be calculated by the Lanczos algorithm \cite{Dagotto94}. The Green's function of the isolated cluster is calculated as
\begin{eqnarray}
G^{C}_{(i,\sigma),(j,\sigma^{\prime})}(z)&=&\langle\Omega|c_{(i,\sigma)}\frac{1}{z-\mathcal{H}_{+}+E_0}c_{(j,\sigma^{\prime})}^{+}|\Omega\rangle \nonumber \\&+&\langle\Omega|c_{(j,\sigma^{\prime})}^{+}\frac{1}{z+\mathcal{H}_{-}-E_0}c_{(i,\sigma)}|\Omega\rangle
\end{eqnarray}
where $\mathcal{H}_{+(-)}$ is the matrix form of the interacting Hamiltonian in the Hilbert space of half pulse(minus) one filling, $E_0$ is the ground state energy level of $\mathcal{H}$. The self-energy of the isolated cluster is defined as $\Sigma_{C}(z)=z-H^{C}_{free}-[G^{C}(z)]^{-1}$, with $H^{C}_{free}$ being the non-interacting Hamiltonian of the isolated cluster.

\begin{figure}[tbp]
\scalebox{0.33}{\includegraphics{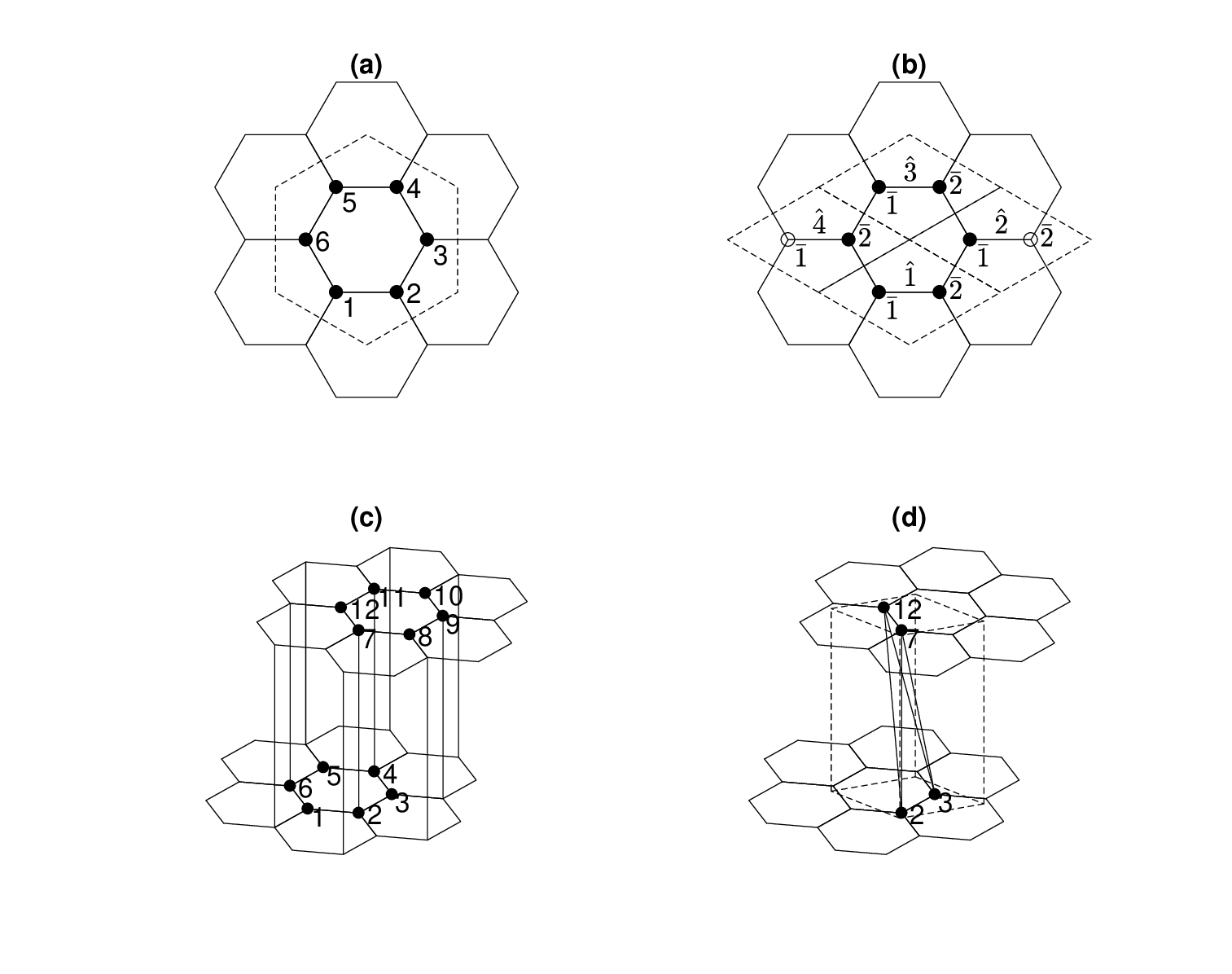}}
\caption{ For single layer honeycomb lattice model, the lattice of clusters in (a) and the lattice of unit cells in (b). The indices of lattice sites within the cluster is presented in figure (a). The indices of unit cells and lattice sites within each unit cells are presented in figure (b). For bilayer honeycomb lattice model, the lattice of clusters in (c) and the lattice of unit cells in (d). }
\label{fig_Cluster1}
\end{figure}

(iii) Thirdly, the self-energy within the primitive unit cell of the honeycomb lattice is obtained by periodization of $\Sigma_{C}(z)$ \cite{QingXiao15}. The lattice site index in the cluster is split into composite index as $i=(\alpha,\bar{i})$ and $j=(\beta,\bar{j})$, where $\alpha$($\beta$) is the index of primitive unit cell, and $\bar{i}$($\bar{j}$) is the index of the lattice site within the primitive unit cell. For example, a six-site cluster and primitive unit cells in single layer honeycomb lattice are marked by dashed lines in Fig. \ref{fig_Cluster1}(a) and (b), respectively. The index $i$($j$) takes the value $\{1,2,3,4,5,6\}$ as shown in Fig. \ref{fig_Cluster1}(a); the index $\bar{i}$($\bar{j}$) takes the value $\{\bar{1},\bar{2}\}$, the index $\alpha$($\beta$) takes the value $\{\hat{1},\hat{2},\hat{3},\hat{4}\}$ as shown in Fig. \ref{fig_Cluster1}(b). The periodization of the self-energy of the isolated cluster into the self-energy within one primitive unit cell is obtained as
\begin{equation}
\Sigma^{P}_{\gamma;(\bar{i},\sigma),(\bar{j},\sigma^{\prime})}(z)\equiv\frac{1}{N_{c}}\sum_{\mathbf{R}_{\beta}-\mathbf{R}_{\alpha}=\mathbf{R}_{\gamma}}\Sigma^{C}_{(\alpha,\bar{i},\sigma),(\beta,\bar{j},\sigma^{\prime})}(z)
\label{equationSigmaP}
\end{equation}
\begin{equation}
\Sigma^{P0}_{(\bar{i},\sigma),(\bar{j},\sigma^{\prime})}(\mathbf{k},z)=\sum_{\gamma}\Sigma^{P}_{\gamma;(\bar{i},\sigma),(\bar{j},\sigma^{\prime})}(z)e^{i\mathbf{k}\cdot\mathbf{R}_{\gamma}}
\end{equation}
where $\mathbf{R}_{\alpha(\beta)}$ is the lattice vector of the primitive unit cell. The cluster might not be divided into a set of complete primitive unit cells. If the lattice site $(\alpha,\bar{i})$ is beyond the cluster's lattice sites (the empty dots in Fig. \ref{fig_Cluster1}(b)), the corresponding matrix elements $\Sigma^{C}_{(\alpha,\bar{i},\sigma),(\beta,\bar{j},\sigma^{\prime})}(z)$ is set to zero in Eq. (\ref{equationSigmaP}). Because there are two lattice sites in one primitive unit cell, $\Sigma^{P0}(\mathbf{k},z)$ only contains on-site and nearest-neighbor terms. Similar scheme of cluster for bilayer honeycomb lattice and the index of lattice sites are shown in Fig. \ref{fig_Cluster1}(c). The lattice sites in the primitive unit cell are chosen as shown in Fig. \ref{fig_Cluster1}(d), so that $\Sigma^{P0}(\mathbf{k},z)$ contains non-diagonal terms between nearest neighbor intra-layer and inter-layer lattice sites.

(iv) At last, the Bloch periodic boundary condition is applied to the nearest-neighbor terms of $\Sigma^{P0}$, which gives the non-diagonal terms in $\Sigma(\mathbf{k},z)$. The diagonal terms in $\Sigma(\mathbf{k},z)$ is the same as those in $\Sigma^{P0}(\mathbf{k},z)$.

The third and fourth step restore the translation symmetry of the original honeycomb lattice. Inserting the self-energy in Eq. (\ref{greenfunction}) gives the Green's function. As comparison, mean field(MF) approximation captures only the on-site electrostatic interaction. The self-energy is assumed to be diagonal and independent of $\mathbf{k}$ and $\varepsilon$, i.e. the self-energy is self-consistently defined as $\Sigma^{MF}_{i\sigma,i\sigma}:=U\langle n_{i\bar{\sigma}}\rangle$.

\begin{figure}[tbp]
\scalebox{0.58}{\includegraphics{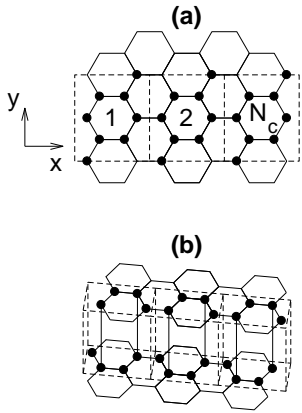}}
\caption{ The scheme of cluster in (a) and (b) for zigzag nanoribbons of single layer and bilayer honeycomb lattice, respectively. }
\label{fig_Cluster3}
\end{figure}

The spectral function of zigzag nanoribbons is also calculated by CPT. The clusters of nanoribbons of single layer and bilayer honeycomb lattices are shown in Fig. \ref{fig_Cluster3}(a) and (b), respectively. The unit cell of nanoribbon contain too many lattice sites to perform ED. Thus, the unit cell is separated into $N_{c}$ clusters along $x$ direction, each containing eight lattice sites. The Green's function of each cluster, $G_{C}^{i}$ with $i\in\{1,2,\cdots,N_{c}\}$, are calculated by ED. The Green's function of the isolated unit cell is obtained by inversion of the block tri-diagonal matric
\begin{eqnarray}
&&G_{C}(z)=\\
&&\left[\begin{array}{cccc}
[G_{C}^{1}(z)]^{-1} & V_{x}^{1,2} & 0 & 0  \\
V_{x}^{2,1} & [G_{C}^{2}(z)]^{-1} & V_{x}^{2,3} & 0  \\
0 & \ddots & \ddots & \ddots  \\
0 & 0  & V_{x}^{N_{c},N_{c}-1} & [G_{C}^{N_{c}}(z)]^{-1} \\
\end{array}\right]^{-1}\nonumber
\end{eqnarray}
where the nondiagonal blocks $V_{x}^{i,i\pm1}$ represent hopping along $x$ direction between neighboring clusters. The Green's function of the nanoribbon is obtained by periodization of the isolated unit cell's Green's function
\begin{equation}
G(k_{y},z)=\frac{G_{C}(z)}{I-(V_{y+}e^{ik_{y}L_{y}}+V_{y-}e^{-ik_{y}L_{y}})G_{C}(z)}
\end{equation}
where $V_{y\pm}$ represent hopping along $\pm y$ direction between unit cells, and $L_{y}$ is the period along $y$ direction.

The solution of CPT preserves the symmetry of the original Hamiltonian. Thus, in the absence of an (anti)ferromagnetic exchange field, the solution does not have (anti)ferromagnetic order. By contrast, the solution of MF could spontaneously break the symmetry and indicate antiferromagnetic order for pristine graphene. In bulk, spontaneous antiferromagnetic order occurs for $U>2.2t$. A more accurate calculation with quantum Monte Carlo shows that the phase transition occur at $U=4.5t$ \cite{Sorella92}. In this article, $U=1.6t$ is used for realistic graphene, allowing the spontaneous antiferromagnetic order to be neglected. For zigzag nanoribbon, the spontaneous antiferromagnetic order could occur with any $U$ \cite{Rossier08,Jung09,Lado14}. The spontaneous antiferromagnetic order is spatially localized at the open boundaries and induces a band gap that is strongly dependent on the width of the nanoribbon. If the width is larger than $8$ nm, the gap becomes negligible. For the nanoribbons in this article, the width is much greater than $8$ nm, so that the spontaneous antiferromagnetic order and the induced band gap could be neglected. As a result, CPT calculation could capture the major physics of the interacting systems in this article.

\begin{acknowledgments}
The project is supported by the National Natural Science Foundation of China (Grant:
11704419). We would like to thank Prof Han-Qing Wu for discussion.
\end{acknowledgments}

\clearpage

\end{document}